# Generative Discovery of Novel Chemical Designs using Diffusion Modeling and Transformer Deep Neural Networks with Application to Deep Eutectic Solvents

Rachel K. Luu[1,2], Marcin Wysokowski[1,3], Markus J. Buehler[1,4]*

[1] Laboratory for Atomistic and Molecular Mechanics (LAMM), Massachusetts Institute of Technology, 77 Massachusetts Ave., Cambridge, MA 02139, USA

[2] Department of Materials Science and Engineering, Massachusetts Institute of Technology, 77 Massachusetts Ave., Cambridge, MA 02139, USA

[3] Faculty of Chemical Technology, Poznan University of Technology, Berdychowo 4, 60965 Poznan, Poland

[4] Center for Computational Science and Engineering, Schwarzman College of Computing, Massachusetts Institute of Technology, 77 Massachusetts Ave., Cambridge, MA 02139, USA

*mbuehler@MIT.EDU

**ABSTRACT**: We report a series of deep learning models to solve complex forward and inverse design problems in molecular modeling and design. Using both diffusion models inspired by nonequilibrium thermodynamics and attention-based transformer architectures, we demonstrate a flexible framework to capture complex chemical structures. First trained on the QM9 dataset and a series of quantum mechanical properties (e.g. homo, lumo, free energy, heat capacity, etc.), we then generalize the model to study and design key properties of deep eutectic solvents. In addition to separate forward and inverse models, we also report an integrated fully prompt-based multi-task generative pretrained transformer model that solves multiple forward, inverse design, and prediction tasks, flexibly and within one model. We show that the multi-task generative model has the overall best performance and allows for flexible integration of multiple objectives, within one model, and for distinct chemistries, suggesting that synergies emerge during training of this large language model. Trained jointly in tasks related to the QM9 dataset and deep eutectic solvents (DESs), the model can predict various quantum mechanical properties and critical properties to achieve deep eutectic solvent behavior. Several novel combinations of DESs are proposed based on this framework.

***Significance statement:*** *We explore how deep learning models can be used to solve complex forward and inverse design problems in molecular modeling and design. Our work shows that a multi-task autoregressive transformer model, trained against a set of complex forward and inverse tasks, can effectively learn from disparate data and make accurate predictions, outperforming conventional approaches by exploiting synergies across datasets and tasks. These results suggest emergent capabilities of large language models for modeling physical phenomena.*

Generative chemistry is an emerging frontier in materials discovery and has been applied to proteins[1–4], organic molecules, inorganics, drug design[5], bioactive materials[6], solid-state materials[7], and architected materials,[8,9] among others. **Figure 1a** shows an overview of the approach implemented, generating molecular structures from chemical building blocks, atoms. Three distinct neural network architectures are used here, a diffusion model with self-/cross-attention (**Figure 1b**), and two transformer architectures (**Figure 1c**). A variety of tasks are implemented, broadly grouped into forward predictions, **Figure 1d** (take a chemical structure written in Simplified Molecular-Input Line-Entry System (SMILES)[10,11], predict its properties) and inverse design tasks, **Figure 1e** (take design conditions, predict candidate SMILES molecular structures). While the diffusion models and the transformer models are each trained separately for each task, the generative pretrained transformer model is fully prompt-based and capable of solving multiple tasks in one model (see **Table S1** for an overview, and **Figure S1** for sample results from the design processes). Details on the models, training and inference process, and other key information, is included in **Supplementary Materials, S1. Materials and Methods.** Here we summarize the key components of the three architectures.

The ***diffusion model*** (**Figure 1b**) implements a thermodynamics-inspired denoising process by which a noisy starting signal is transformed into the solution via a deep neural network U-net architecture[12–14]. The U-net architecture used in the diffusion model features 1D convolutional layers mixed with self-/cross-attention layers. The convolutional layers capture hierarchical patterns, and self-attention captures long-range relationships in the signal. The denoising process is conditioned using cross-attention mechanisms on a set of parameters. This results in an iterative procedure, $z_{i-1} = z_i - \varepsilon_i'$, where denoising happens by calculating the noise to be removed, $\varepsilon_i'$, using the deep neural network p (which defines a reverse Markov chain operator), where

$$\varepsilon_i' = p(z_i, C_i; t_i) \qquad (1)$$

with $z_i$ being the noisy signal at step $i$, $C_i$ the conditioning used, and $t_i$ the time step). The diffusion model predicts tokens as one-hot encoding, from which we then sample the token with the highest probability after denoising is complete.

The ***attention-based transformer model*** (**Figure 1c, left**) is implemented in several variants[15,16]. Model T1 is an autoregressive decoder-only architecture that produces solutions iteratively from a start token during inference and using cross-attention with the conditioning features. The key mathematical operation is the masked attention mechanism[15,16], defined as

$$\text{Attention}(Q, K, V; M) = \text{softmax}\left(\frac{QK^T + M}{\sqrt{d_k}}\right) V \qquad (2)$$

Where Q, K, and V are inputs to the attention layer (all the same in self-attention and Q=input, and K=V=conditioning in cross-attention). In the inverse model, causal masking using a triangular masking matrix *M* is used in the self-attention step so that the model can only attend to tokens to previous tokens. We use gumbel softmax sampling during inference, which allows us to tune the creativity of the model (a certain level of noise, defined by the sampling temperature *T*, is added to the predicted logit distributions, from which we then sample the predicted token)[17,18]. In the forward transformer model, an encoder-only strategy is used to relate the input (tokenized and embedded molecular structure) to the output, with self-attention, realizing model T1' (**Figure S7**).

In the multi-task transformer model T2 (**Figure 1c, right**), no cross-attention is necessary. All conditioning and distinction of various tasks is provided directly by the input prompt (which is first fed into the model, and the model then continues the sequence to provide the answer). All code is developed in PyTorch[19] and training is performed using an Adam optimizer[20].

We use two datasets (details on tokenization etc., see **Supplementary Materials *S1.1 Tokenizer and Datasets***). The first is Quantum Machines 9 (QM9), a dataset[21] of 133,885 molecules, including their SMILES text encodings and their quantum chemical properties. Second, a smaller dataset curated for this study, is used to capture properties of deep eutectic solvents (DESs). Deep eutectic solvents (DESs) are innovative mixtures that are characterized by their significantly lower melting points relative to those of their constituents[22–24]. These solvents provide a clean and sustainable medium for the processing and synthesis of advanced materials[22] and are considered as impactful solvent strategies[25,26].

In the following, we present results produced by these models. While the diffusion model and transformer model T1 take a certain type of input (tokenized SMILES strings or numerical values of target properties), the multi-task transformer model T2 is entirely based on text input, and computes solutions by providing prompts (overview of some of the prompts trained for, see **Table 1**). The purpose of utilizing these 3 distinct architectures and a total of five trained models is to assess the overall best strategy to dealing with the problem at hand. We specifically hypothesize that using a multi-task integrated model T2 that can be trained simultaneously on diverse datasets, and multiple tasks, can yield certain synergies and perform better overall.

This formulation of the multi-task integrated transformer model T2 offers a much more flexible approach to various kinds of design and analysis problems. The input in the transformer model T2 is purely text, both for numerical input and output as well as SMILES codes. In this framework, we use an input:

```
~Calculate<CC1=CC2CC2CC1O>
```

which is then transformed by T2 into:

```
~Calculate<CC1=CC2CC2CC1O> /-0.889,-0.176,0.230,0.020,-0.278,-
0.342,0.300,0.370,-0.028,-0.028,-0.028,-0.028|$
```

Where the output of the model is highlighted in **bold**. The resulting numbers of this calculation task to obtain the 12 QM9 properties can then be converted from string format into floating point numbers for further analysis.

**Figure 2** shows a comparison of design objectives labeled as Ground Truth (GT) versus predicted values (Prediction), for the mechanical properties captured in the QM9 dataset. In this analysis, we use the workflow shown in **Figure 2a**; where this analysis tests *both* the forward and inverse tasks simultaneously. Results are shown for the three architectures used, the diffusion model (**Figure 2b**, R2=0.92), the transformer models T1/T1' (**Figure 2c**, R2=0.94) and the prompt-based transformer model T2 (**Figure 2d**, R2=0.97). To complement these results, **Figure S2** shows the forward model performance alone, predicted values over GT. **Figure S2a** shows the results for the diffusion model (R2=0.97) and **Figure S2b** the results for the transformer model T1 (R2=0.96). **Figure S2c** depicts results for the transformer model T2 (R2=0.99), for the QM9 dataset.

We now analyze a few sample structures generated by each of the three approaches (**Figure 3)** and provide an in-depth discussion of the results and performance. In the following analysis, only novel chemical designs are considered that do not exist in the training or validation set. **Figure 3a** shows results based on the diffusion model, **Figure 3b** for the transformer model T1, and **Figure 3c** for the multi-task integrated transformer model T2, each for different conditioning parameters plotting GT versus prediction using the regression model. All three models have a strong capacity to discover novel structures (included in neither training or test set). We find that the multi-task integrated transformer model T2 produces generally a higher fraction of novel molecular designs than the other two models. This, combined with the better overall performance with respect to forward and inverse tasks, and the overall greater flexibility, indicates broader advantages of this architecture over the other two.

The multi-task integrated transformer model T2 is text based and can carry out multiple tasks. For example, in the generative task, an input:

```
~Generate<-0.767,-0.274,0.284,-0.020,-0.332,-0.386,0.128,0.235,-0.124,-0.124,-
0.124,-0.124>,
```

leads to the output (highlighted in **bold**):

```
~Generate<-0.767,-0.274,0.284,-0.020,-0.332,-0.386,0.128,0.235,-0.124,-0.124,-
0.124,-0.124> /CC1NCC2C1NC2=O|$
```

The previous examples considered *de novo* design tasks. However, in some cases we wish to either start with an existing chemical design or solve a partial design task where we only want to redesign part of a molecule. Such tasks can be addressed quite well using generative models, especially the diffusion approach. **Figure 4a-c** shows structural discovery experiments using inpainting strategy, using the inverse diffusion model. **Figure 4a-b** show results; where the first three SMILES characters are given as a fixed constraint and the rest as initial solution that can change. **Figure 4c** shows the generation results for an unconstrained design, but with an initial guess (same as in panel **Figure 4a**). The highest R2 score between the desired properties and the predicted properties is obtained for the case in panel **c** (R2=0.86), the second highest for the case in panel a (R2=0.85) and the worst for the results in **Figure 4b** (R2=0.82). Overall, the generative method discovers molecules that are close to the target, but the best result is obtained for the unconstrained case shown in **Figure 4c**, which makes intuitive sense. The structures in **Figure 4a**-b are novel, whereas the structure in **Figure 4c** is seen in the dataset. This experiment shows how by using inpainting and masking we can direct the model towards discovery of new molecules that meet a specific target and interpolate between different levels of novelty.

**Figure 4d-f** show similar structural discovery experiments using the autoregressive transformer model. A distinction to the diffusion model is that due to the autoregressive nature it does not allow for inpainting experiments; and hence, we use only three initial symbols in the SMILES string to initiate generation (these are provided to the model after the start token). The model completes these initial design ideas and produces

molecules that meet the design demand well. The R2 values for the results in panels **Figure 4d-f** are R2=0.86, R2=0.82, and R2=0.82, respectively.

We now focus on the most complex set of tasks, making forward and inverse predictions for deep eutectic solvents (DESs). As an emerging class of mixtures, discovery of new combinations of hydrogen bond acceptors (HBAs) and donors (HBDs) that achieve DESs behavior is rather expensive and time consuming in the laboratory setting. Therefore, there is a growing emphasis in using computational design and machine learning algorithms as tools to support DESs discovery and predict their features[27], including density[28], viscosity[29,30] and surface tension[31]. Melting temperature plays an essential role in the design of new deep eutectic solvents [32,33]. Therefore, we introduce a new dataset of DESs that contains 402 different DES compositions, where HBAs and HBDs are represented by SMILES. The dataset consists of the melting temperature of individual HBAs and HBDs, as well as melting points of DESs mixtures in relation to their HBAs:HBDs mole ratio (see **Supplementary Material**). The new dataset for this class of materials is much smaller than the QM9 dataset (402 data points vs. 130,000+). When training the models solely on the small DES dataset, we expectedly find lower performance potentially due to influential outliers and overfitting (**Figure S3** depicts performance for the forward task **Figure S3a** for the diffusion model (R2=0.59) and **Figure S3b** transformer model T1' (R2=0.55). Since the forward model does not perform well, we do not consider models trained for the inverse design task.

We propose a training strategy that can deal with such complex tasks while still providing an avenue to integrate the two datasets, both DES and QM9, for learned synergies between the various problems. Using the multi-task integrated transformer model T2, we train it against a variety of tasks (see **Table 1**). **Figure 5** shows the results obtained using the integrated multi-task transformer model T2 applied to design DES molecular pairs with associated properties, trained against a combined QM9-based set of tasks (predict properties and design molecules) and DES tasks (calculate $T_{melt}$ of individual DES components, $T_{DESmelt}$ for pairs of DES components, as well as the ratio to achieve desired $T_{DESmelt}$).

**Figure 5a** and **b** depicts the model performance regarding DES specific prediction tasks, respectively $T_{DESmelt}$ and ratio (R2=0.93) from a pair of molecules, and $T_{melt}$, the melting temperature of individual components (R2=0.86). While the predictions do not reach the same level as for the QM9 tasks shown in **Figure 2d**, the results are encouraging for machine-learning assisted DESs development. Next, we test model performance for the generative design task. **Figure 5c** and **d** show two example results for the design task, revealing two novel DES designs. In **Figure 5c**, $T_{DESmelt}$ and the desired mole ratio is provided, and a pair of molecules are predicted (resulting in monoethylcholine chloride and 4-methylcatechol). In **Figure 5d**, $T_{DESmelt}$ is provided, and both the pair of molecules and the mole ratio are predicted (resulting in benzyltrimethylammonium chloride and 2-aminopropane-1,2,3-tricarboxylic acid). A sample input is:

```
~GenerateDES_withratio<-0.314>
```

which leads to

```
~GenerateDES_withratio<-0.314>
/C[N+](C)(C)CC1=CC=CC=C1.[Cl-],C(C(=O)O)C(CC(=O)O)(C(=O)O)N,0.013|$
```

The prediction is a combination of two SMILES strings and a floating-point number that describes the mole ratio.

The model was able to identify the functional groups responsible for accepting and donating the hydrogen bond and proposed new DESs composed of, for instance, monoethylcholine chloride (hydrogen bond donor count of 2) and 4-methylcatechol (hydrogen bond donor count=2). The investigation of choline-based compounds and diols, along with aromatic alcohols[34,35], has been extensively documented in scientific literature. The model proposed a combination of benzyltrimethylammonium chloride ($T_{melt}$ 239°C) with 2-aminopropane-1,2,3-tricarboxylic acid ($T_{melt}$ 156°C) in molar ratio 1:1 that will result in deep depression of melting point to 33.3°C **(Supporting Table S2)**. A combination of quaternary ammonium salts and carboxylic acids has been also explored in literature[36]. For completeness, we show that the model can also rediscover already-known designs and accurately calculate DES properties for known DESs composed of tetrabutylammonium bromide and aspartic acid (**Figure 5e**) and N,N-Diethylethanolammonium chloride and glycerol (**Figure 5f**). It is encouraging that the model can rediscover known DES compositions. Since the dataset is extremely small, further validation of these results is needed.

The model has shown to make similar decisions as human experts in the field of DES. The results generated by the model can inspire researchers during the design phase of DES, motivating them to experimentally prepare suggested combinations of molecules and assess their $T_{melt}$ at varying mole ratios. Subsequently, the additional experimental data could be leveraged to enhance the model's performance by adding new data to the training set.

We presented a flexible platform for materials discovery using frameworks of diffusion models and transformer architectures (**Figure 1b-c**). We can easily incorporate these models into a range of applications, and the use of distinct architectures offers flexible avenues. The diffusion model can easily solve inpainting problems (**Figure 4**). All generative models can solve degenerate design tasks and suggest multiple candidate solutions for a given objective. The transformer models generally perform well, and the use of autoregressive approaches with multi-task training in the style of generative pretrained transformer models[37] provides the overall best performance, and highest level of flexibility (**Figure 2c**). The multi-task model works exceptionally well for the QM9 dataset, both for forward and inverse design tasks (**Figures 2-4**) and can also be applied to a new class of chemistry and associated new set of tasks (**Figure 5**).

DESs are an emerging class of designer solvents in modern and sustainable chemistry for which relatively little data exists. By utilizing our newly developed dataset and applying it to our multi-task transformer model T2, we have demonstrated the impressive abilities in expediting the exploration and examination of DES properties and design. Despite using a small dataset, the model can predict diverse properties and generate new DESs compositions, like monoethylcholine halide and 4-methylcatechol. Considering the importance of DESs in toxic gas absorption, energy storage, metal extraction; future avenues of study include the design of the DESs with multiple tailored properties like conductivity, absorption capacity, viscosity, surface tension.

The generative models show high potential in being applied to a large variety of tasks to accelerate discovery and materials design. Another key insight is that the superb performance of the multi-task transformer model T2 has not only the best performance overall but can also be integrated efficiently with the smaller DES dataset to still yield reasonable performance. It outperforms the separately trained forward and inverse models, suggesting emergent capabilities in the language models, in general language contexts and including modeling physical phenomena[16,38–41]. This is an exciting development and offers promising future research opportunities for other systems.

**Authors' Contributions:** M.J.B. developed the overall concept and the algorithms, designed the various deep learning models, developed the codes, oversaw the work, trained and tested the models, and drafted the paper. R.K.L. trained and tested ML models and helped analyze results, reviewed and edited the original draft, and helped with curating the datasets. M.W. conceptualized the deep eutectic solvent work, analyzed results, reviewed, and edited the original draft, and curated the deep eutectic solvent dataset.

**Data and code availability**: The code and data are available at: https://github.com/lamm-mit/MoleculeDiffusionTransformer. Additional data, including the DES dataset, is included as part of the Supplementary Information.

**Conflict of Interest Statement:** The authors have no conflicts to disclose.

**Acknowledgements**: This material is based upon work supported by the National Science Foundation Graduate Research Fellowship under Grant No. 2141064. The authors acknowledge support from the Army Research Office (W911NF1920098 & W911NF2220213), as well as USDA (2021-69012-35978). M.W. acknowledges The Kosciuszko Foundation and National Science Center – Poland (SONATA 17 2021/43/D/ST5/00853) for supporting his research stay in the USA.

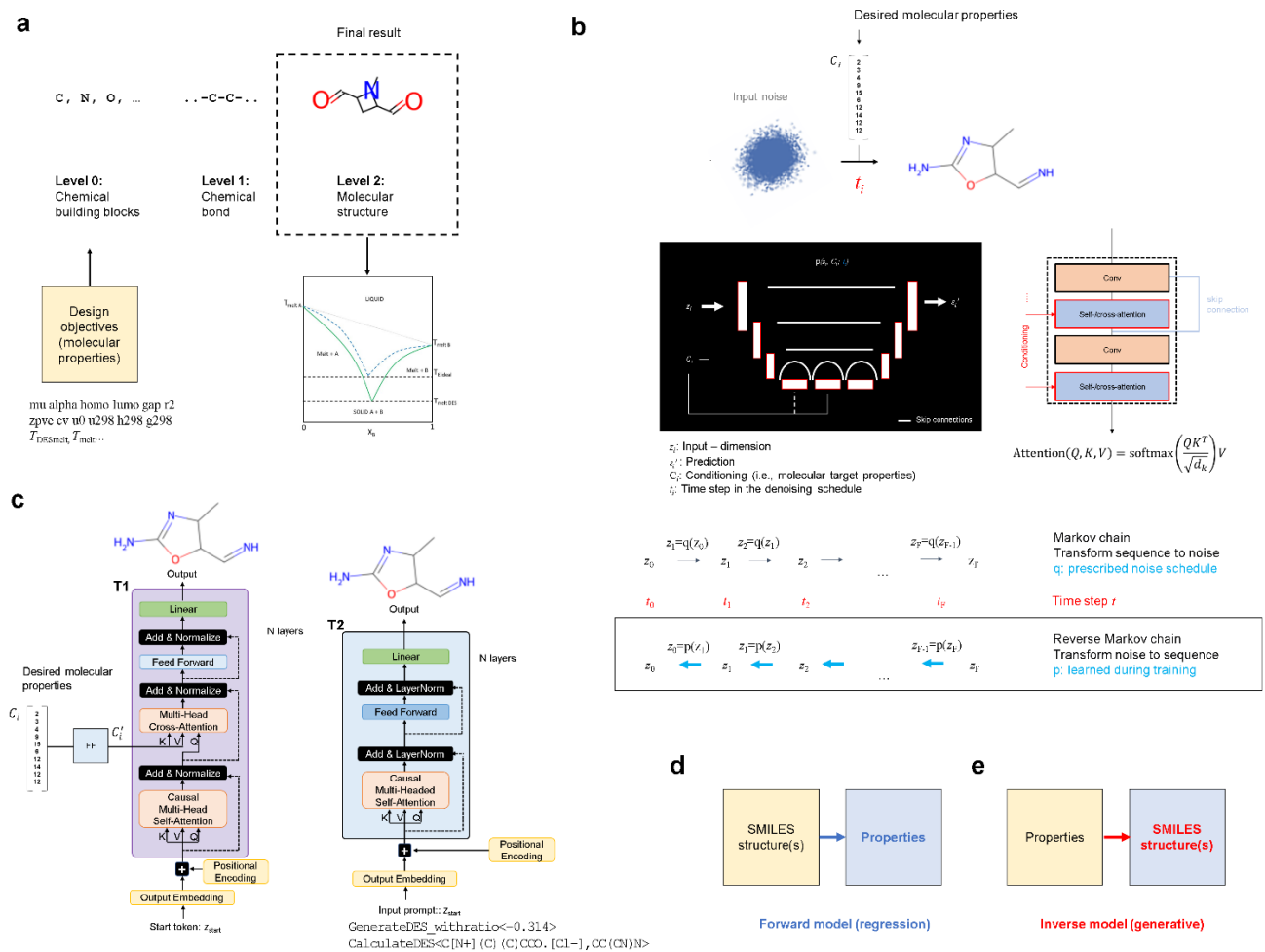

**Figure 1**: Approach for generating molecular structures using generative deep learning **a**, overview of the approach used, including sample design objectives. **b**, diffusion model, **c** the autoregressive transformer model, realized in several versions (T1, autoregressive transformer model where the design objectives are implemented via cross-attention, T1' (see **Figure S7**), and T2, an autoregressive transformer model where tasks are implemented via various input prompts). Panels **d** and **e** show the tasks solved using the models.

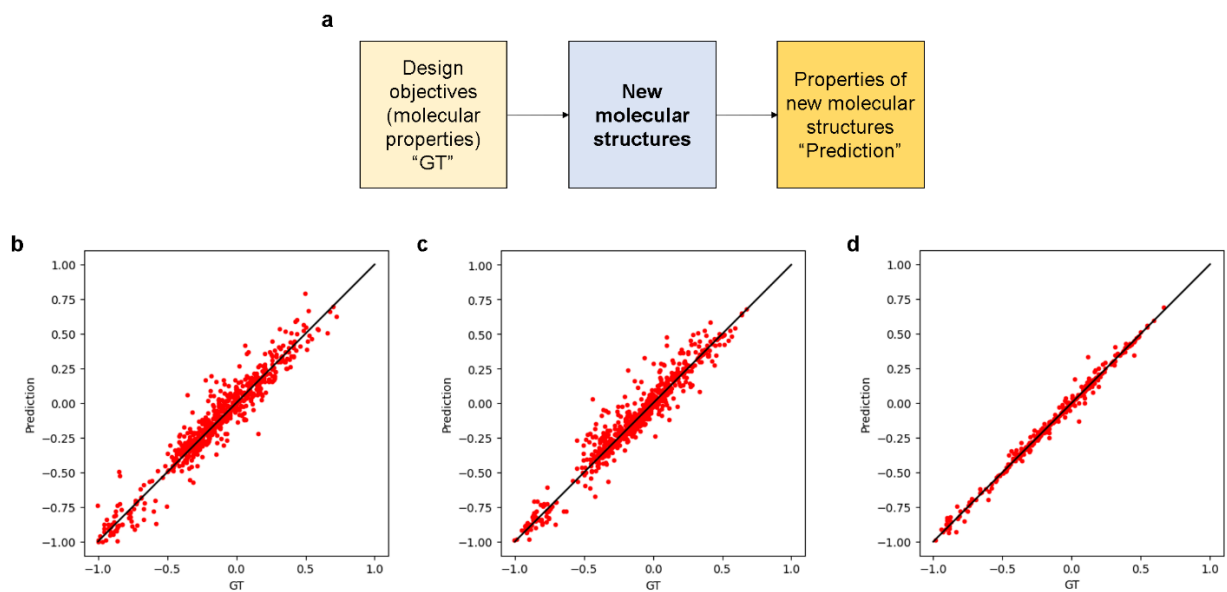

**Figure 2**: Comparing design objectives, Ground Truth versus Prediction, in the workflow as shown in **a** (this analysis tests both forward/inverse tasks) Results are shown for three architectures used, the diffusion model (**b**, $R2=0.92$), the transformer model T1/T1' (**c**, $R2=0.94$) and the multi-task prompt-based transformer model T2 (**d**, $R2=0.97$).

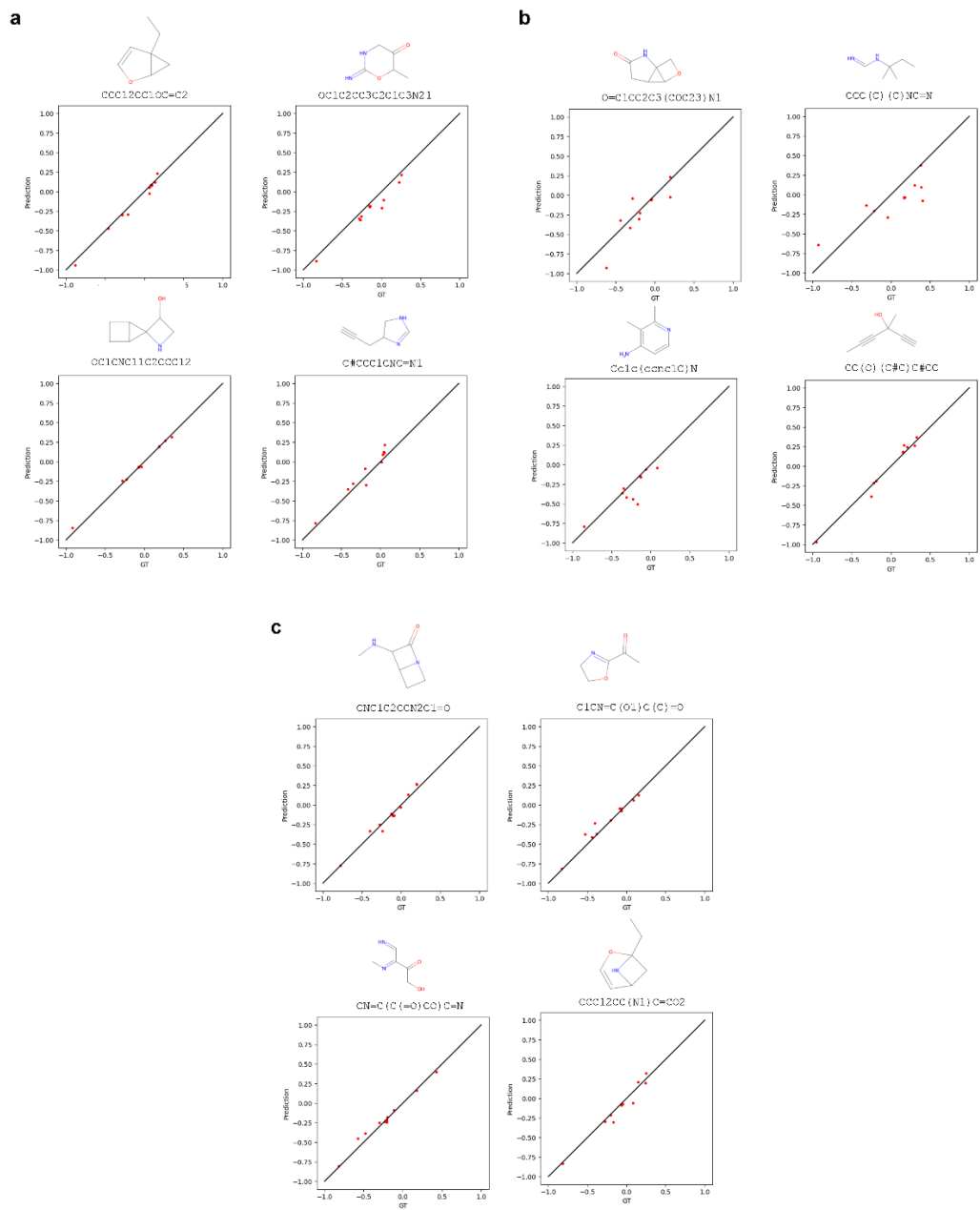

**Figure 3**: Novel sample structures generated by the diffusion model (**a**), transformer model T1 (**b**), and transformer model T2 (**c**), for different conditioning parameters, plotting Ground Truth versus Prediction using the regression model.

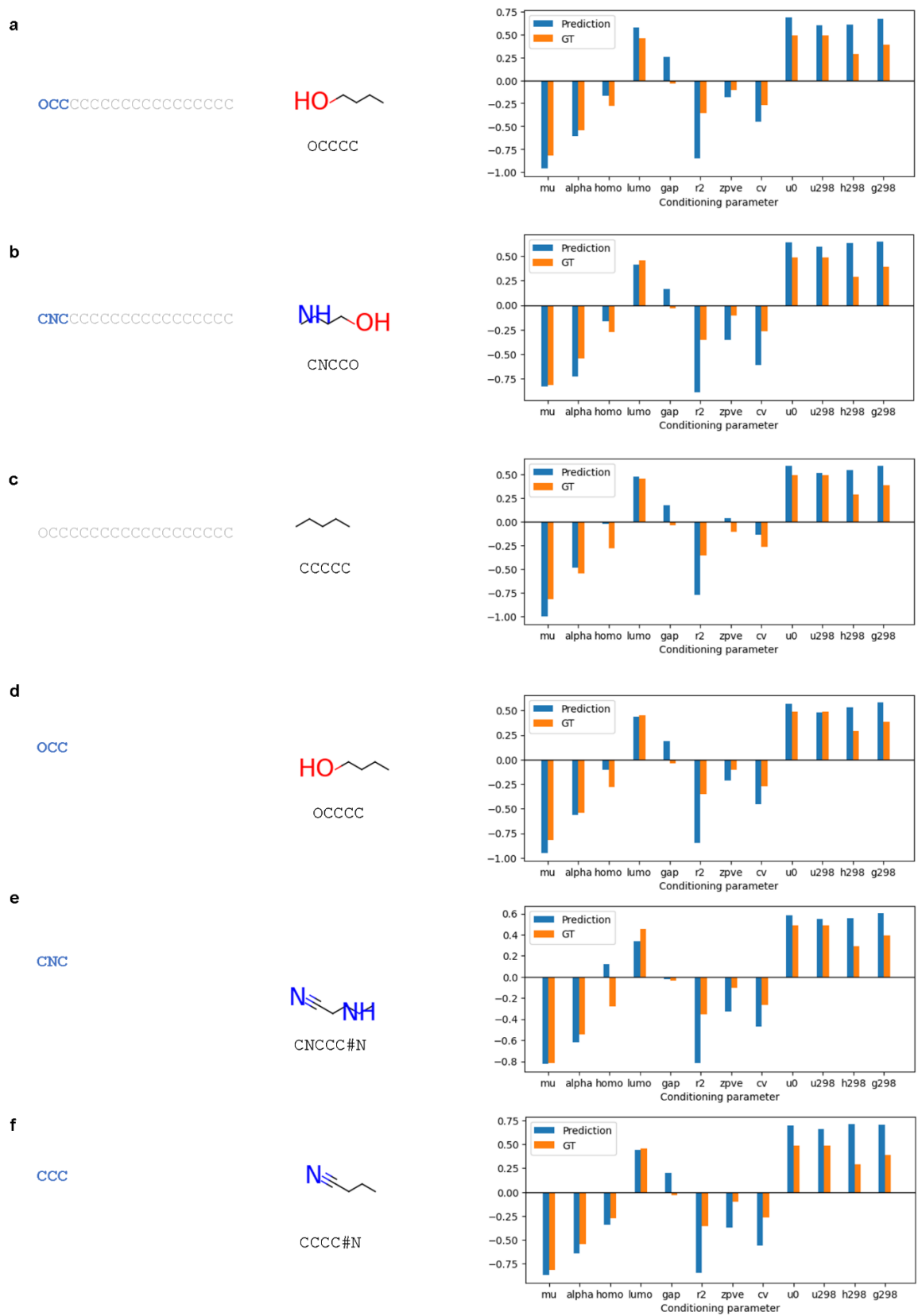

**Figure 4**: Structural discovery experiments using inpainting strategy with inverse diffusion model (**a-c**). All samples are generated using 25 sampling steps, 2 resampling steps, and conditioning scale=2.0 to increase 'creativity' of the model. **A-b** show results based on an initial structural guess, where the first three SMILES characters printed in **bold blue** are given as a fixed constraint (the other SMILES characters, all "C", in grey color are also provided to the model as initial guess but are changeable as the model discovers the solution. **c,** generation results for a completely unconstrained design, but with an initial guess (same as in **a**). **d-f** show similar experiments using the autoregressive transformer model.

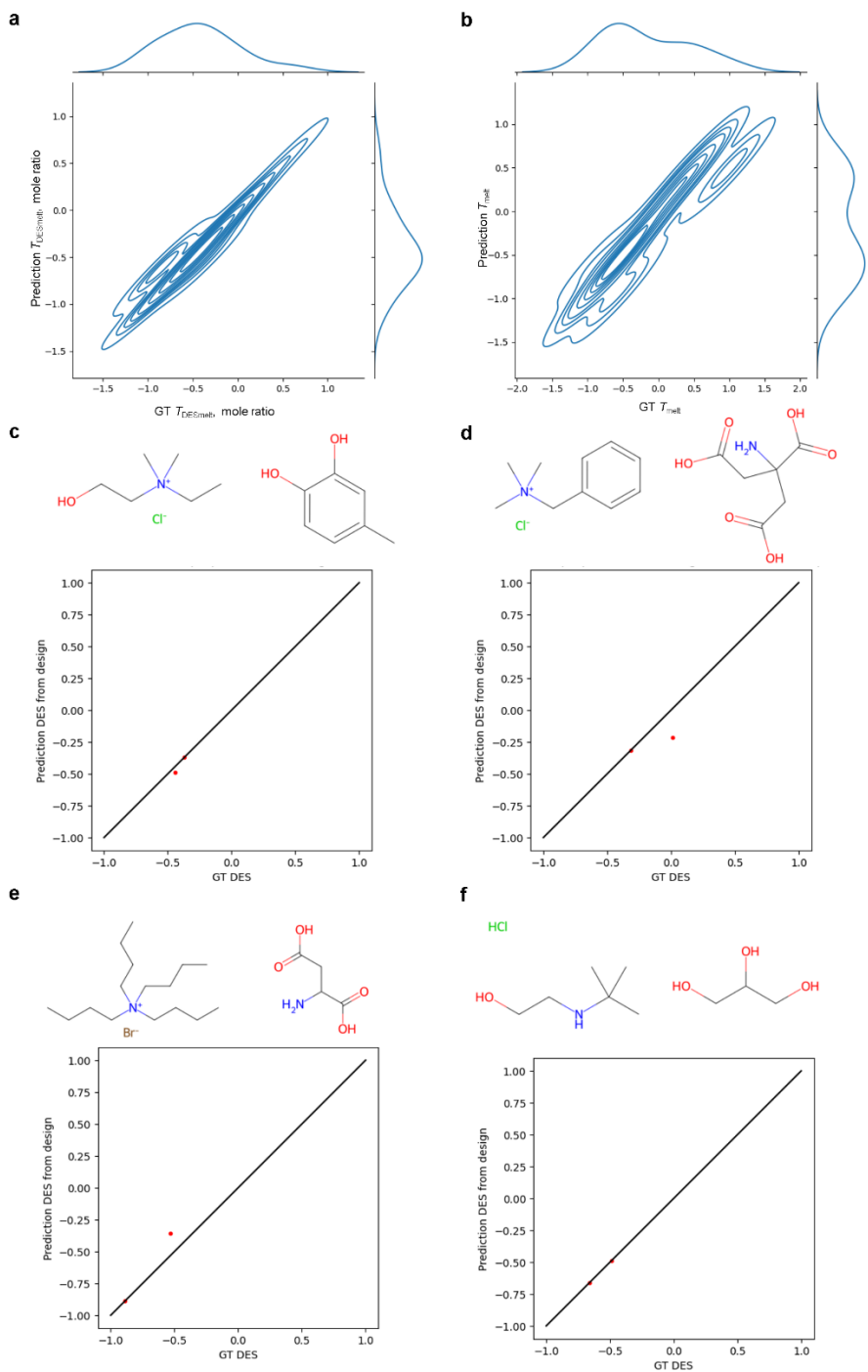

**Figure 5**: Integrated multi-task transformer model T2 applied to design DESs molecular pairs and associated properties. Panels **a** and **b** show the model performance, Ground Truth versus Prediction, trained on two different prediction tasks: **a** $T_{DESmelt}$ and mole ratio ($R2=0.93$), **b** $T_{melt}$, of the individual components ($R2=0.86$). Panels **c** and **d** show two examples for results from design tasks (additional results, see **Supporting Table S2**). **e** and **f** document the discovery of existing designs.

**Table 1:** Sample prompts trained for in the multi-task transformer model T2. Additional sets of tokens are used to encapsulate various tasks and input/output boundaries (~: start token, <..> encapsulate task, as well as /..| to encapsulate prediction, and $ as end token).

| Prompt | Prediction |
|---|---|
| `~Calculate<CC1=CC2CC2CC1O>` | Calculate QM9 properties from SMILES input |
| `~Generate<-0.767,-0.274,0.284,-0.020,-0.332,-0.386,0.128,0.235,-0.124,-0.124,-0.124>` | Design a molecule, expressed as a SMILES output, to meet the target QM9 properties |
| `~GenerateDES<-0.551,-0.570>` | Generate a pair of DES molecules that meet the mole ratio and $T_{\text{DESmelt}}$ target |
| `~GenerateDES_withratio<-0.487>` | Generate a pair of DES molecules and associated mole ratio that meet the $T_{\text{DESmelt}}$ target |
| `~CalculateTmelt<C[P+](C1=CC=CC=C1)(C2=CC=CC=C2)C3=CC=CC=C3.[Br-]>` | Calculate $T_{\text{melt}}$ for a DES component |

# Generative Discovery of Novel Chemical Designs using Diffusion Modeling and Transformer Deep Neural Networks with Application to Deep Eutectic Solvents

Rachel K. Luu[1,2], Marcin Wysokowski[1], Markus J. Buehler[1,3]*

[1] Laboratory for Atomistic and Molecular Mechanics (LAMM), Massachusetts Institute of Technology, 77 Massachusetts Ave., Cambridge, MA 02139, USA
[2] Department of Materials Science and Engineering, Massachusetts Institute of Technology, 77 Massachusetts Ave., Cambridge, MA 02139, USA
[3] Center for Computational Science and Engineering, Schwarzman College of Computing, Massachusetts Institute of Technology, 77 Massachusetts Ave., Cambridge, MA 02139, USA
*mbuehler@MIT.EDU

## SUPPLEMENTARY INFORMATION

## S1. Materials and Methods

### S1.1 Tokenizer and Datasets

Our generative models work with the Simplified Molecular-Input Line-Entry System (SMILES) language to describe chemical structures [1]. SMILES is a form of textual representation that use grammatical rules to encode information about the bonds and atoms of a molecule. This representation allows for complex structural chemistry to be described in simple 1D text encodings and is a highly reliable tool for generative chemistry [2,3]. Text input is then transformed into a numerical representation using a character-level tokenizer, trained for each of the datasets. **Figure S4** depicts pair correlations and data distribution in the QM9 dataset used for training.

**Figures S5** shows an example summary of the tokenizer for the diffusion model (a) and the transformer model (b). The transformer model includes start "~" and end "$" tokens to mark the beginning and end of a sequence. For transformer model T2, additional tokens are introduced to mark the beginning "<" and end ">" of the task, as well as the beginning "/" and end "|" of the prediction.

For training and testing, we use the Quantum Machines 9 (QM9) dataset which is a quantum chemistry dataset consisting of 133,885 molecules. The dataset is composed of the molecules' SMILES codes and its 12 associated quantum mechanical properties: Dipole moment, polarizability, highest occupied molecular orbital, lowest unoccupied molecular orbital, energy gap, expectation value <$R^2$>, zero point energy, internal energy, internal energy at 298.15K, enthalpy at 298.15K, v at 298.15K, and heat capacity at 298.15K.

The DES dataset was collected from peer reviewed published papers in the field[4-41]. DES are classified as having a lower melting temperature as a mixture than its individual constituents. The dataset is composed of 402 DES combinations, each solvent separated by their HBAs and HBDs SMILES, the respective $T_{melt}$ of both individual components, the ratio of each component described as a normalized mole fraction ($x_{HBA} + x_{HBD} = 1$), as well as the deep eutectic melting temperature $T_{DESmelt}$ of the mixture related to the HBA:HBD ratio.

All numerical data presented in this work is normalized between -1 and 1 prior to training.

We anticipate that the DES tasks can be explored further in future work, especially considering additional/alternative properties. This challenging task will require integration of the scientific community working on deep eutectic solvents and creation of a large, unified database that will include the DES SMILES codes with the target properties. Additionally, reporting of the solid-liquid equilibrium (SLE) diagrams should be standard practice in the context of experimental discovery of new DESs. This approach not only facilitates effortless data gathering, but also reinforces model training, thereby amplifying its advantages.

### S1.2 Diffusion model

**Figure 2c** visualizes the denoising process, where the top defines a Markov chain operator q that adds Gaussian noise step-by-step (according to a defined noise schedule that defines how much noise $\varepsilon_i$ is added at each step $i$), translating the physical solution (either the stress-deformation response or the encoded image) $z_0$ (left) into pure noise, $z_F$ (right) following:

$$z_{i+1} = q(z_i) \tag{S1}$$

The deep neural network is trained to *reverse* this process. This is done by identifying an operator p that maximizes the likelihood of the training data. Once trained, it provides a means to translate noise to solutions and thereby realizing the transition illustrated in **Figure 1b** (noise to solution), in a step-by-step fashion as indicated in the lower row of this panel:

$$z_{i-1} = p(z_i) \tag{S2}$$

We use the L2 distance to define the loss, measuring the error of the actual added noise $\varepsilon_i$ and the predicted added noise $\varepsilon_i'$. Hence, the trained diffusion model can predict the added noise. Knowing this quantity then allows us to realize a numerical solution to the denoising problem and hence be used to generate the next iteration of the denoised sequence:

$$z_{i-1} = z_i - \varepsilon_i' \tag{S3}$$

Therein the sequence $z_i$ at step $i$ is transformed by removing the noise $\varepsilon_i'$ whereas this process is performed iteratively. Thereby, the neural network predicts, given the current state $z_i$, the noise to be removed at a given time step $t_i$ in the denoising process (area marked with a rectangle). The conditioning signal $C_i$ is encoded using additive Fourier positional encoding following the approach suggested in[42].

For cross-attention, the conditioning signal is concatenated with the time signal so that $C_{i,\text{total}} = [C_i, t_i]$. This provides us with the opportunity to train a single neural network to learn denoising steps, while jointly considering the conditioning and the time step $t_i$ in the reverse process. The U-net architecture features 1D convolutional layers mixed with self-/cross-attention layers. The convolutional layers capture hierarchical patterns, and the attention layers learn long-range relationships between patterns. This type of architecture can effectively capture the complex mechanisms such as the process of molecular construction[43]. We use the noise schedule, sampling and training processes proposed in proposed in[44] since it yields computationally efficient sampling strategies, obtaining results within 96 denoising steps. **Table S2** provides additional details about the model architecture parameters.

The functioning of both models in forward and inverse directions is schematically shown in **Figure 1b.**

*Inpainting tasks*

Inpainting tasks are solved by providing an initial guess of the structure, on which denoising is then performed. We use a mask to identify tokens that are not allowed to change, such as shown in the examples depicted in **Figure 4a-c**.

*S1.3 Attention based transformer models*

**Figure 1c** depicts a summary of the transformer architectures used here (left: transformer T1, right: transformer T2). Model T1 represents an autoregressive decoder-only architecture that produces solutions iteratively from a start token during inference and using cross-attention with the conditioning features $C_i'$, where $C_i'$ is the output of a feed-forward layer that expands the dimension from [number of input/conditioning features, 1] to [number of input/conditioning features, $d_C$]. The key mathematical operation is the masked attention mechanism[45,46], defined as

$$\text{Attention}(Q, K, V; M) = \text{softmax}\left(\frac{QK^T + M}{\sqrt{d_k}}\right)V \tag{S4}$$

The attention calculation is implemented in multi-headed form by using parallelly stacked attention layers. Instead of only computing the attention once, in the multi-head strategy we divide the input into segments (in the dimension of the embedding) and then computes the scaled dot-product attention over each segment in parallel, allowing the model to jointly attend to information from different representation subspaces at different positions:

$$\text{MultiHead}(Q, K, V) = \text{Concat}(\text{head}_1, \ldots, \text{head}_h)W^O \tag{S5a}$$
$$\text{head}_i = \text{Attention}(QW_i^Q, KW_i^K, VW_i^V) \tag{S5b}$$

where the projections are parameter matrices $W_i^Q \in \mathbb{R}^{d_{model} \times d_q}$, $W_i^K \in \mathbb{R}^{d_{model} \times d_k}$, $W_i^V \in \mathbb{R}^{d_{model} \times d_v}$ and $W^O \in \mathbb{R}^{hd_v \times d_{model}}$. In self-attention, all *Q, K, V* come from either input or output embeddings (or other sources) only, cross-attention calculations here are performed with *Q* from the conditional embeddings and *K, V* from the encoded desired molecular properties, computed from the input embedding via a fully connected feed forward network from the *n*-dimensional feature vector $C_i$. During training, the target represents samples, conditioned by the associated molecular properties $C_i$ whereas the model is trained to predict the output. Causal masking using a triangular masking matrix *M* is used in the self-attention step so that the model can only attend to tokens to the left (*i.e.*, previous tokens). The forward transformer model is a variant T1' consisting of an encoder-model only, as shown in **Figure S7**.

As shown in **Figure 1c**, a start token $\mathcal{T}$ is added at the beginning of the sequence and an end token $\mathcal{E}$ at the end, so that

$$z = [\mathcal{T}, z_1, z_2, \ldots, z_N, \mathcal{E}] \tag{S6}$$

During generation, the start token $\mathcal{T}$ is first fed into the model and the output is predicted from it. During sampling iterations, this process is repeated until the full output is produced. We use gumbel softmax sampling [47,48] during inference, which allows us to tune the creativity of the model (a certain level of noise, defined by the sampling temperature *T*, is added to the predicted logit distributions, from which we then sample the predicted token). While this can be desirable in generative tasks to add more variations in the predictions (*T* around or larger than 1), forward prediction tasks are best conducted using low sampling temperatures (*T*=0.1 or lower). **Figure S6 depicts the** effect of sampling temperature on QM9 predictions (**Figure S6 a**, argmax reflecting *T*=0, **Figure S6b-d**, gumbel softmax sampling [47,48] with *T*=1.5, 1., and 0.1, respectively. This shows that the results of the forward task are stable in spite of higher sampling temperatures.

In the multi-task transformer model T2, no cross-attention is necessary. All conditioning and distinction of various tasks is provided by the input prompt. Additional sets of tokens are defined, as exemplified in **Table 1**, to encapsulate various tasks and input/output boundaries (~: start token $\mathcal{T}$, <..> encapsulate task, /..| to encapsulate prediction, and $ as end token $\mathcal{E}$).

*S1.4 Training process, generative strategy, and other hyperparameters*

All code is developed in PyTorch, for further details see[49]. All machine learning training is performed using an Adam optimizer[50], with a learning rate of 0.0002. Some generative tasks, specifically for the smaller DES dataset, produce a larger variability of solutions. One strategy to narrow in on higher performing candidates is to check whether predicted designs satisfy the required properties (akin to the framework shown in **Figure 2**). A threshold can be defined, e.g. a MSE error of <0.2, to decide which design to accept or reject. Similarly, some designs are solutions that exist already in the dataset. While it can be a useful exercise to include these as well, we focus on accepted designs that are novel for the presentation of the generative results shown in this paper. The only exception are the results in **Figure 5e-f** where we display molecular structures that have already been known. The purpose of this exercise was to show, for the DES case, that the model can discover existing, known structures. The same holds for the QM9 design tasks.

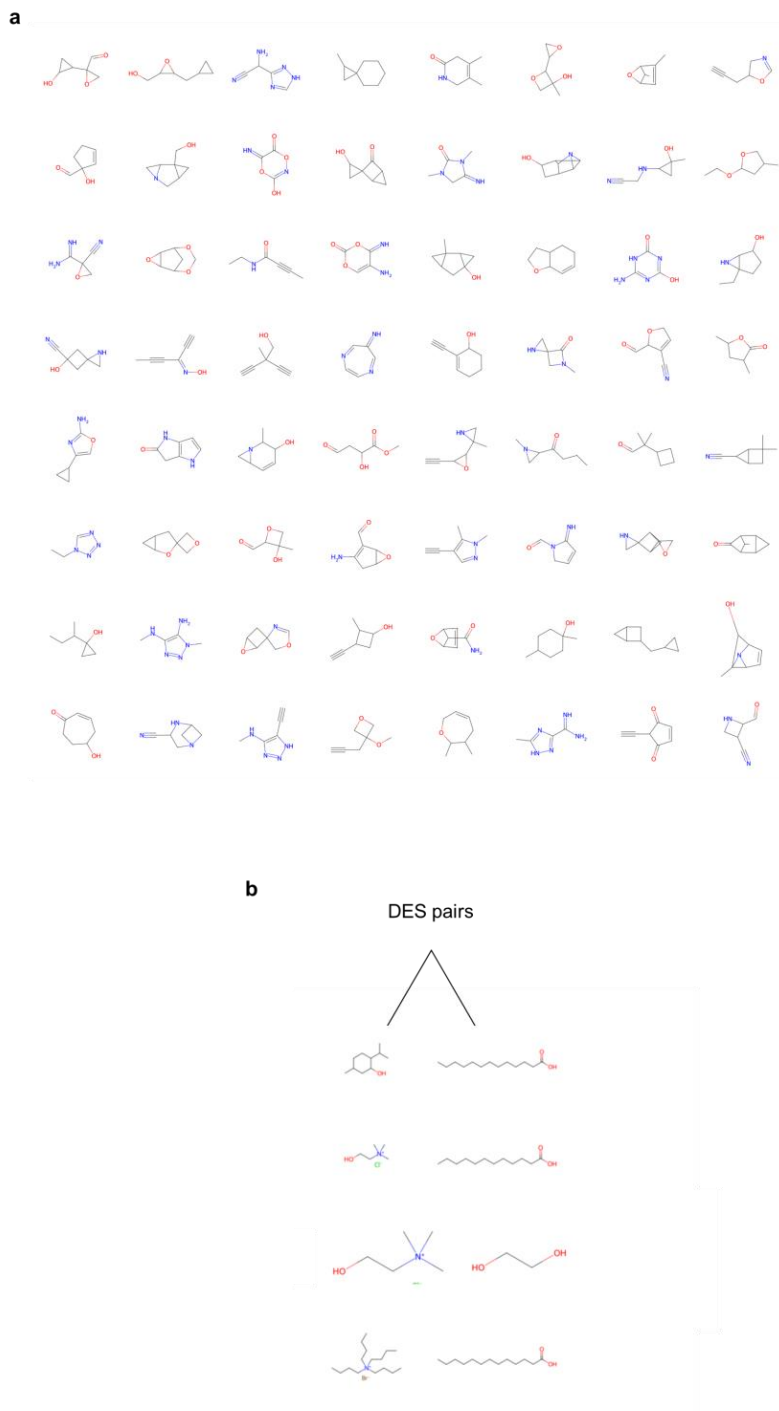

**Figure S1**: A set of 64 molecules generated using the diffusion model (**a**, showing sample molecules to meet quantum mechanical properties as featured in the QM9 dataset, and **b**, deep eutectic solvent designs that feature pairs of molecules).

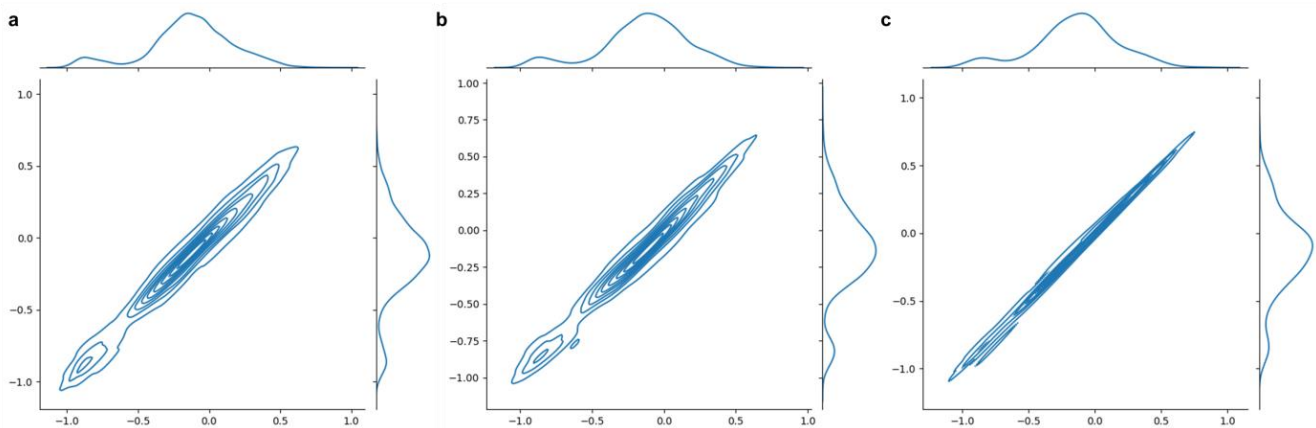

**Figure S2**: Forward model performance, predicted values over ground truth. Panel **a** shows the results for the forward diffusion model (R2=0.97), **b** the results for the forward transformer model T1' (R2=0.96) and **c** the results for the multi-task transformer model T2 (R2=0.99), for the QM9 dataset.

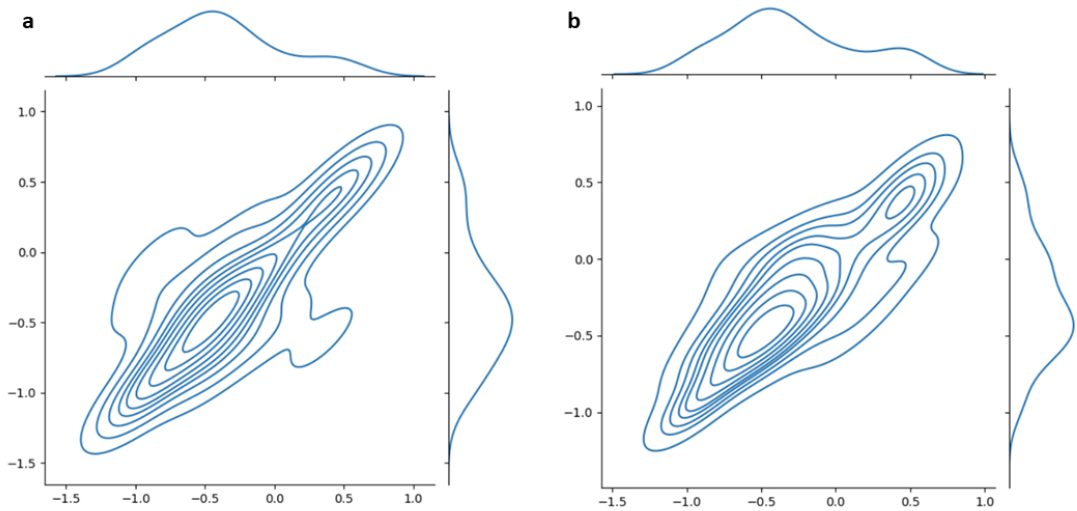

**Figure S3**: Forward model performance results for training on only 402 values contained in the DES dataset, predicted values over ground truth. Panel **a** shows the results for the forward diffusion model (R2=0.59), **b** the results for the transformer model T1' (R2=0.55).

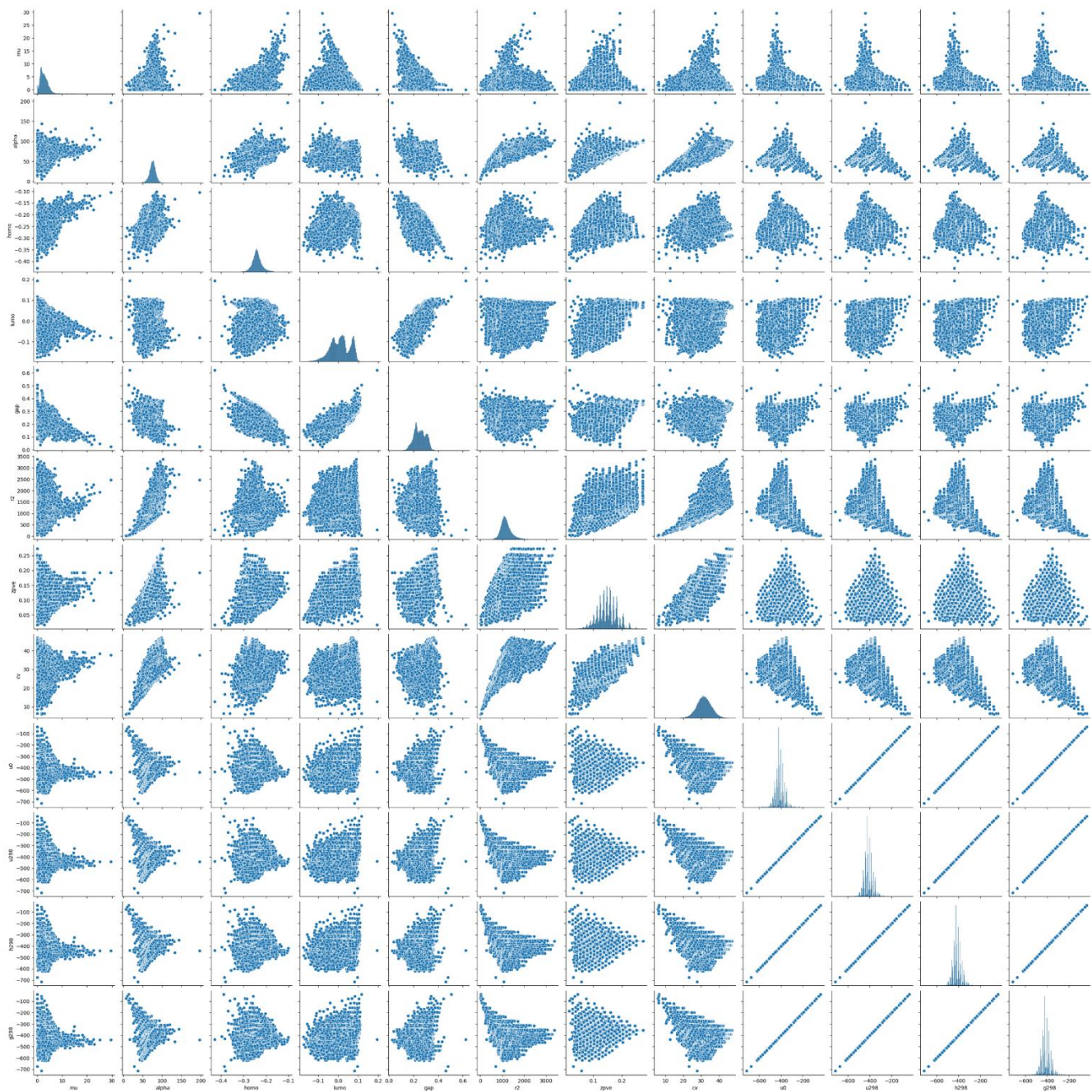

**Figure S4**: Pair correlations and data distribution in the QM9 dataset used for training, across the 12 properties used.

**a**

```
DICTIONARY y_data
{'num_words': None, 'filters': '', 'lower':
False, 'split': ' ', 'char_level': True,
'oov_token': None, 'document_count': 133885,
'word_counts': '{"C": 787482, "N": 109306, "O":
181161, "#": 37027, "=": 106335, "(": 126532,
")": 126532, "1": 270020, "c": 59074, "[": 9918,
"n": 30458, "H": 8247, "]": 9918, "o": 6836,
"3": 34756, "+": 1705, "-": 1808, "2": 121675,
"F": 3314, "4": 5186, "5": 242}', 'word_docs':
'{"C": 131656, "N": 74300, "O": 111215, "#":
32432, "=": 76005, ")": 92249, "(": 92249, "1":
119887, "H": 7941, "n": 16002, "[": 7941, "]":
7941, "c": 17698, "o": 6546, "+": 1705, "3":
17627, "-": 1808, "2": 60295, "F": 2163, "4":
2593, "5": 121}', 'index_docs': '{"1": 131656,
"7": 74300, "3": 111215, "10": 32432, "8":
76005, "5": 92249, "4": 92249, "2": 119887,
"15": 7941, "12": 16002, "13": 7941, "14": 7941,
"9": 17698, "16": 6546, "20": 1705, "11": 17627,
"19": 1808, "6": 60295, "18": 2163, "17": 2593,
"21": 121}', 'index_word': '{"1": "C", "2": "1",
"3": "O", "4": "(", "5": ")", "6": "2", "7":
"N", "8": "=", "9": "c", "10": "#", "11": "3",
"12": "n", "13": "[", "14": "]", "15": "H",
"16": "o", "17": "4", "18": "F", "19": "-",
"20": "+", "21": "5"}', 'word_index': '{"C": 1,
"1": 2, "O": 3, "(": 4, ")": 5, "2": 6, "N": 7,
"=": 8, "c": 9, "#": 10, "3": 11, "n": 12, "[":
13, "]": 14, "H": 15, "o": 16, "4": 17, "F": 18,
"-": 19, "+": 20, "5": 21}'}
```

**b**

```
DICTIONARY y_data
{'num_words': None, 'filters': '', 'lower':
False, 'split': ' ', 'char_level': True,
'oov_token': None, 'document_count': 133885,
'word_counts': '{"@": 133885, "C": 787482, "$":
133885, "N": 109306, "O": 181161, "#": 37027,
"=": 106335, "(": 126532, ")": 126532, "1":
270020, "c": 59074, "[": 9918, "n": 30458, "H":
8247, "]": 9918, "o": 6836, "3": 34756, "+":
1705, "-": 1808, "2": 121675, "F": 3314, "4":
5186, "5": 242}', 'word_docs': '{"$": 133885,
"@": 133885, "C": 131656, "N": 74300, "O":
111215, "#": 32432, "=": 76005, ")": 92249, "(":
92249, "1": 119887, "n": 16002, "H": 7941, "c":
17698, "]": 7941, "[": 7941, "o": 6546, "3":
17627, "+": 1705, "-": 1808, "2": 60295, "F":
2163, "4": 2593, "5": 121}', 'index_docs':
'{"5": 133885, "4": 133885, "1": 131656, "9":
74300, "3": 111215, "12": 32432, "10": 76005,
"7": 92249, "6": 92249, "2": 119887, "14":
16002, "17": 7941, "11": 17698, "16": 7941,
"15": 7941, "18": 6546, "13": 17627, "22": 1705,
"21": 1808, "8": 60295, "20": 2163, "19": 2593,
"23": 121}', 'index_word': '{"1": "C", "2": "1",
"3": "O", "4": "@", "5": "$", "6": "(", "7":
")", "8": "2", "9": "N", "10": "=", "11": "c",
"12": "#", "13": "3", "14": "n", "15": "[",
"16": "]", "17": "H", "18": "o", "19": "4",
"20": "F", "21": "-", "22": "+", "23": "5"}',
'word_index': '{"C": 1, "1": 2, "O": 3, "@": 4,
"$": 5, "(": 6, ")": 7, "2": 8, "N": 9, "=": 10,
"c": 11, "#": 12, "3": 13, "n": 14, "[": 15,
"]": 16, "H": 17, "o": 18, "4": 19, "F": 20, "-
": 21, "+": 22, "5": 23}'}
```

**Figures S5**: Example summary of the tokenizer for the diffusion model (a) and the transformer model (b). The transformer model includes start ('@') and end ('$') tokens to mark the beginning and end of a sequence.

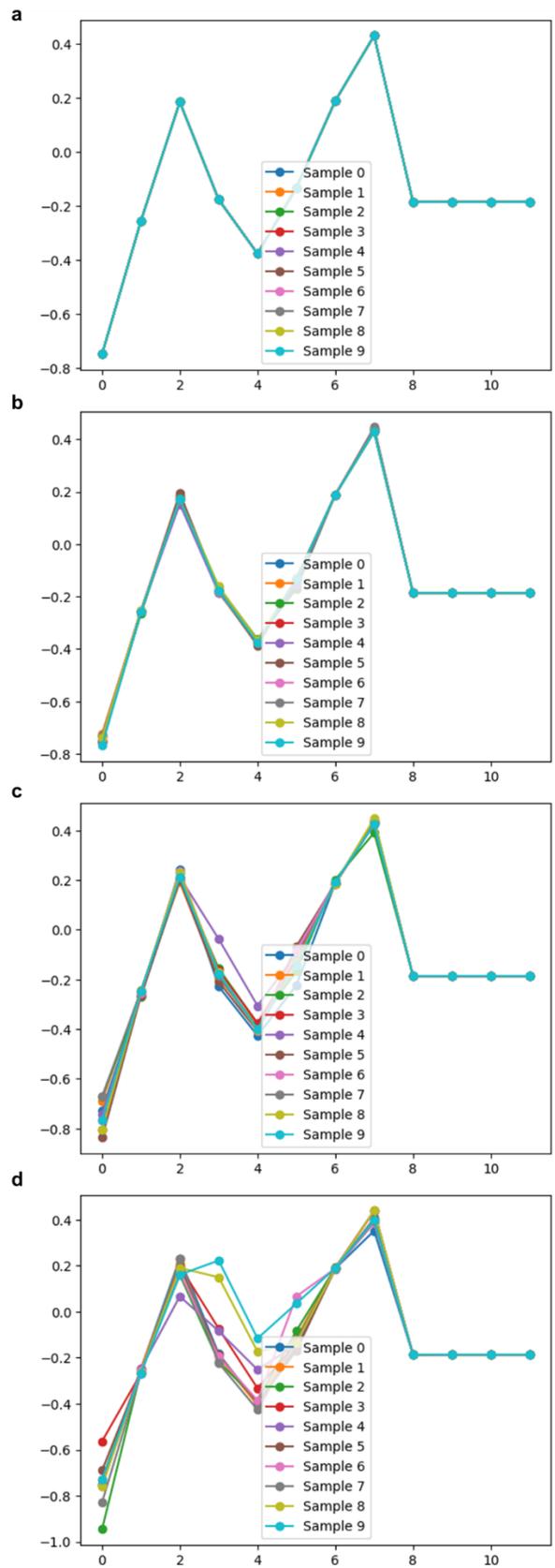

**Figure S6:** Effect of sampling temperature on QM9 predictions (**a**, argmax, **b-d**, gumbel softmax sampling with $T$=1.5, 1., and 0.1, respectively. This shows that the results of the forward task are stable.

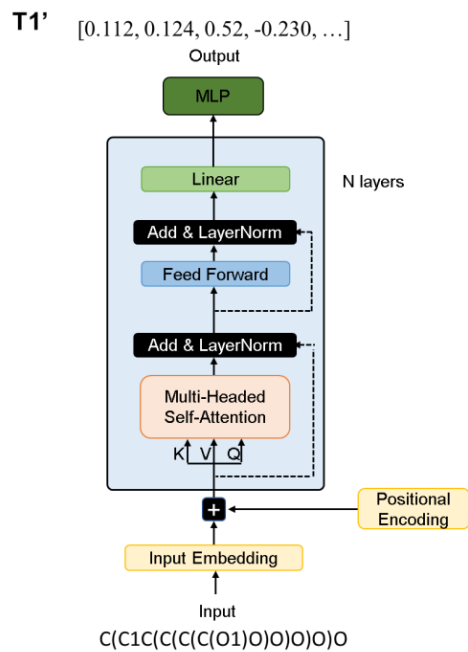

**Figure S7:** Visualization of the forward model T1', an encoder-only model that predicts molecular properties from input sequences.

**Table S1:** Parameters used in the models.

| Model | Key parameters | Tasks trained for | Notes |
|---|---|---|---|
| Diffusion model, forward | Channels=64, classifier-free guidance, conditioning and Fourier positional embedding dimension=64, 2 attention layers and 3 ResNet blocks in each U-net layer | Predict properties from SMILES input | Trained on QM9 data and DES data separately |
| Diffusion model, inverse | Channels=128, classifier-free guidance, conditioning and Fourier positional embedding dimension=64, 4 attention layers and 3 ResNet blocks in each U-net layer, 2 self-attention layers before downsampling | Generate new molecular designs as SMILES, based on target properties | Trained on QM9 data |
| Transformer model T1', forward | Channels=256, 16 heads, depth=6, feed forward multiplier=S2 (512 channels), embedding dimension=8; encoder-only setup where self-attention is used on the input signal | Predict properties from SMILES input | Trained on QM9 data and DES data separately |
| Transformer model T1, inverse | Channels=256, 8 heads, depth=6, feed forward multiplier =4 (1024 channels), embedding dimension=8, cross-attention embedding dimension=8; decoder only | Generate new molecular designs as SMILES, based on target properties | Trained on QM9 data |
| Transformer model T2 | Channels=256, 16 heads, depth=18, feed forward multiplier =2 (512 channels), embedding dimension=8, cross-attention embedding dimension=8; decoder-only | Solve various calculation tasks for QM9 properties, DES properties, as well as generative tasks to design new molecules and DESs | Jointly trained on QM9 data and DES data |

**Table S2:** Various designed DES combinations, all novel structures, obtained using the multi-task integrated transformer model T2.

| # | SMILES_HBA | SMILES_HBD | X_HBA (ratio) | Mixed_Tmelt (degrees C) |
|---|---|---|---|---|
| 0 | CC[N+](C)(C)CCO.[Cl-] | CC1=CC=C(C(=C1)O)O | 0.329905 | 10.02104 |
| 1 | CCNC[N+](CCC)(CCC)CCC.[Br-] | C(C(=O)O)O | 0.249805 | -9.49672 |
| 2 | CCCCCCCCCC(=O)O | CCCCCCCCCCCC(=O)O | 0.60002 | 23.2136 |
| 3 | C[N+](C)(C)CC1=CC=CC=C1.[Cl-] | C(C(=O)O)C(CC(=O)O)(C(=O)O)N | 0.500785 | 33.33392 |
| 4 | CCCCCCCCCP(=O)(CCCCCCC)CCCCCCC | CC1=CC(=C(C=C1)C(C)C)O | 0.30009 | 21.22568 |
| 5 | C[N+](C)(C)CC1=CC=CC=C1.[Cl-] | CC1=CC=C(C=C1)S(=O)O | 0.499895 | 14.35832 |

## Supplementary References